\documentclass{article}
\usepackage{amsmath,amssymb}

\begin{document}

\title{Nonseparability and simultaneous readability}
\author{Riuji Mochizuki\thanks{E-mail: rjmochi@tdc.ac.jp}\\
Laboratory of Physics, Tokyo Dental College,\\ 2-9-7 Kandasurugadai, Chiyoda-ku, Tokyo 101-0062, Japan }

\maketitle

\begin{abstract}
In this study, we investigate quantum nonseparability between an observed system and a measuring apparatus, or multiple measuring apparatuses.  We show that the physical meaning of the outcome of the measuring apparatus obtained by weak measurement with a post-selection differs critically from that without any post-selection.  In this study, the nonseparability plays the essential role, which is shown to be the same in a simultaneous conventional von Neumann-type measurement of multiple observables.  From this viewpoint, we suggest a new concept, known as {\it simultaneous readability}, which is the possibility that multiple measuring apparatuses will give the proper information of the observed system simultaneously.  Next, we show that different components of the spin of an electron are not simultaneously measurable even if it has EPR correlation with another electron.   
\end{abstract}

%\subjectindex{A60}

%\maketitle
\newpage
\section{Introduction}

Nonseparability is one of the central concepts of quantum mechanics.  The unitary evolution of a quantum system obeys the Schr\"{o}dinger equation but its nonunitary change, i.e., reduction of the wavepacket or its replacement, is also unavoidable in the measurement process.  Nonseparability is produced in unitary evolution and appears in the nonunitary change; thus we need careful consideration even after the interaction between the observed system and the measuring apparatus has ended.  In this paper, we employ d'Espagnat's argument\cite{Esp} which was used to clarify the nonseparability of EPR-correlated pairs\cite{EPR}\cite{Bohm} to study simultaneous measurability from the viewpoint of the quantum measurement theory.  We consider the density matrix of the unified system of the observed system and the measuring apparatuses to clarify their nonseparability.

In the second section, we discuss weak measurement and weak values in view of the nonseparability.  Since their concept was developed\cite{Aha2}\cite{Aha25}, many papers concerning them have been written.   Applications of the weak measurement technique have been studied in some of them.  For example, Lundeen et al.\cite{Lun1} have suggested the direct measurement of wavefunctions, which are experimentally verified\cite{Mal}\cite{Sal}.  On the other hand, more papers have been written based on the interpretation that the weak values should be conditional expectation values or others of a like nature.  In this interpretation, the weak value
\[
\langle\hat A\rangle_{FI}^w\equiv\frac{\langle F|\hat A|I\rangle}{\langle F|I\rangle}
\]
is regarded as the expectation value of $\hat A$ for both an initial state $|I\rangle$ and a final state $|F\rangle$, which would be obtained by post-selection following the weak measurement of $\hat A$ for $|I\rangle$.  Because the outcomes obtained by the weak measurement without any post-selection agree with the ordinary measured values,   it may be expected that the weak values would be interpreted as the expectational values even with the post-selection.   This interpretation seems to be supported by many authors\cite{Aha3}\cite{Kocsis}\cite{Wise1}\cite{Hosoya2}\cite{Aha4} with some differences in their interpretation details.  In our previous paper\cite{Mochi}, though we could not decide against this conjecture, we have noted that  some mistakes and insufficiency exist in some important papers on weak measurement and weak values.  We have also shown that the real part of the weak value provides the back-action of the weak measurement to the post-selection.  In addition, the imaginary part of the weak values has been shown to be interpreted similarly by Dressel et al.\cite{Dre1}.

 What we read in the measurement process are not the observables of the observed system but the outcomes of the measuring apparatuses.  These correspond to those in some cases but not in all cases.  The expectation value of the pointer's position, obtained by the weak measurement with the post-selection, has been verified to correspond to the real part of the weak value.  The weak value obtained without any post-selection has been verified to agree with the expectation value of the observable of the observed system.  However, we have not authenticated whether the readout obtained by the weak measurement with the post-selection corresponds to the expectation value of the observable of the observed system.  In this paper, we show that the post-selection causes the fatal effect on the weak measurement due to the nonseparability between the observed system and the measuring apparatus.  The weak values with the post-selection should not be interpreted as expectation values in general, which differ from the weakly measured values without any post-selection. This conclusion is consistent with the discussion in our previous paper\cite{Mochi}.

 With a similar argument, we consider the simultaneous von Neumann-type measurement\cite{von}.  Though two observables belonging to the Hilbert spaces of different measuring apparatuses commute, it does not guarantee that they give the information of the observed system simultaneously.  We cannot, in general, regard both of the readouts of the measuring apparatuses as the expectation values of the corresponding observables of the observed system because of the nonseparability, though either of them can be regard so.  Thus, we suggest a new concept called {\it simultaneous readability} of the observables of the multiple measuring apparatuses, which is the separability between them.  This is possibility that they will give the proper information of the observed system simultaneously. We show that the simultaneous readability with a certain interaction Hamiltonian is the necessary and sufficient condition of the simultaneous measurability of the corresponding observables of the observed system.  The density matrix of the unified system plays the essential role in our discussion.  It reflects the fact that the nonunitary change of quantum states and the nonseparability are indispensable in the quantum measurement theory\cite{Esp}.

Then, we try to solve two troublesome problems of the simultaneous measurability in the latter sections.   As mentioned in \cite{Heisenberg}, it has not been obvious whether we can know the expectation values of two noncommuting observables simultaneously if we prepare some eigenstate.  For example, it seems possible that we can know the different components of an electron's spin simultaneously if the $x$-component of the spin is measured for the eigenstate of its $z$-component.   Moreover, we can measure the different components of the spins of the EPR-correlated electrons\cite{EPR}\cite{Bohm}, which seemingly enables us to know the different components of the spin of {\it one} electron.  Keeping these discussions in view, Ozawa\cite{Ozawa2}\cite{Ozawa1} discussed the simultaneous measurability in a state-dependent formulation.  We demonstrate by extending the discussion in the preceding sections that the different components of a single electron's spin cannot be observed simultaneously even in such cases.

\section{Weak measurement and weak value}

First, we quickly check that the real part of the weak value agrees with the expectation value $\overline x$ of the pointer's position, obtained by the corresponding weak measurement\cite{Aha2}\cite{Lun1}.  
The interaction Hamiltonian $\hat H_I$ between an observable $\hat A$ of the observed system and the momentum $\hat \pi$ of the pointer is  
\begin{equation}
\hat H_I\equiv g\hat A\hat \pi,\label{eq:Hamiltonian}
\end{equation}
where $g$ is the coupling constant.   $\hat H_I$ is assumed to be constant and roughly equivalent to the total Hamiltonian over some interaction time $t$.   The initial wavefunction $\phi (x)$ of the measuring apparatus is assumed to be
\begin{equation}
\phi (x)=\langle x|\phi\rangle =\Big({1\over \sqrt{2\pi}\sigma}\Big)^{1/2}\exp\Big( -{(x-x_0)^2\over 4\sigma^2}\Big).
\label{eq:wave}
\end{equation}
The initial state $|\Phi(0)\rangle=|I\rangle |\phi\rangle$, where $|I\rangle$ is the initial state of the observed system, of the unified system of the observed system and the measuring apparatus evolves unitarily obeying the Schr\"{o}dinger equation:
\begin{equation}
i\hbar{d\over dt}|\Phi(t)\rangle =\hat H|\Phi (t)\rangle \sim\hat H_I|\Phi (t)\rangle ,
\end{equation}
and becomes
\begin{equation}
|\Phi(t)\rangle =\exp\Big(-\frac{i\hat H_It}{\hbar}\Big)|\Phi(0)\rangle.\label{eq:full}
\end{equation}
Up to the first order of $g$,  
\begin{equation}
|\Phi_1 (t)\rangle =|I\rangle |\phi\rangle -{igt\over\hbar}\hat A|I\rangle\hat\pi |\phi\rangle  .\label{eq:gtichiji}
\end{equation}

Instead, we can equally describe the unified system by means of the density matrix
\begin{equation}
\rho_1(t)=|\Phi_1(t)\rangle\langle\Phi_1(t)|.\label{eq:matrix}
\end{equation}
Without any post-selection, the expectation value $\overline x$ of the pointer's position $\hat x$ for this state is 
\begin{equation}
\begin{split}
\overline x &\equiv{\rm Tr}\big[\rho_1(t)\hat x\big]\\
%&=\langle I|I\rangle\langle\phi |\hat x|\phi\rangle -\frac{igt}{\hbar}\langle I|\hat A|I\rangle\langle\phi |[\hat x, \hat \pi]|\phi\rangle\\
&=x_0+gt\langle I|\hat A|I\rangle.\label{eq:futsuunokitaichi}
\end{split}
\end{equation}
$\overline x$ can be written also in the form
\begin{equation}
\overline x ={\rm Tr}[\rho_1^{(m)}(t)\hat x],
\end{equation}
where $\rho_1^{(m)}(t)$ is the partial density matrix only of the measuring apparatus defined as 
\begin{equation}
\begin{split}
\rho_1^{(m)}(t)&\equiv {\rm Tr}^{(s)}[\rho_1(t)]\\
&=|\phi\rangle\langle\phi |-\frac{igt}{\hbar}\langle I|\hat A|I\rangle \Big(\hat \pi |\phi\rangle\langle\phi |-|\phi\rangle\langle\phi |\hat \pi \Big),\label{eq:rhoo}
\end{split}
\end{equation}
and ${\rm Tr}^{(s)}$ means that the operation of taking the trace is carried over only in the Hilbert space of the observed system.  

On the other hand, with the post-selection $|F\rangle$, the expectation value $\overline x_F$ of the pointer's position is
\begin{equation}
\begin{split}
\overline x_{F} &={\rm Tr}[\rho^{(m)}_{1F}(t)\hat x]\\
&=|\langle F |I\rangle |^2\Big[x_0+gt\Re\langle\hat A\rangle_{FI}^{w}\Big],\label{eq:postari}
\end{split}
\end{equation}
where
\begin{equation}
\rho^{(m)}_{1F}(t)\equiv {\rm Tr}^{(s)}[\rho_1(t)|F\rangle\langle F |].\label{eq:rhoF}
\end{equation}
%and
%\begin{equation}
%\langle\hat A\rangle_{F,I}\equiv\frac{\langle F |\hat A|I\rangle}{\langle F |I\rangle}.\label{eq:teigi}
%\end{equation}
%is the weak value of an operator $\hat A$ for the initial state $|I\rangle$ and the final state $|F\rangle$.

These calculations have shown the following.  Without any post-selection, the readout of the pointer's position corresponds to the expectation value of the observable for the initial state belonging to the observed system.  With the post-selection, the readout of the pointer's position corresponds to the real part of the weak value.  Thus, it remains to be authenticated whether the readout of the pointer's position obtained by the weak measurement with the post-selection reflects the information about the observable for the initial and the final states of the observed system.  Because the partial density matrix (\ref{eq:rhoF}) is obtained after the measurement of the projection operator $\hat F\equiv|F\rangle\langle F|$, our question  is written in another form: Can we read both the expectation values of $\hat A$ and $\hat F$ for the state (\ref{eq:gtichiji}) after the interaction between the observed system and the measuring apparatus has ended?  

To answer this question, we assume that the ensemble $S$ of the observed system and the ensemble $M$ of the measuring apparatus after their interaction are both separately obtained by combining all the elements of subensembles, each of which is described by its own ket.  Then, each element of $S$ belongs to one of the subensembles $E_{i},\ i =1,2,\cdots$ described by $|s_{i}\rangle$ and each element of $M$ belongs to one of the  subensembles $E_{\alpha},\ {\alpha}=1,2,\cdots$ described by $|m_{\alpha}\rangle$, such that the subensemble $\varepsilon_{i,\alpha}$ of the unified system, whose elements belong to both $E_i$ and $E_{\alpha}$, is described by the density matrix 
\begin{equation}
\rho_{i,\alpha}=|s_{i}\rangle |m_{\alpha}\rangle\langle m_{\alpha}|\langle s_{i}|.
\end{equation}

Because the unified system's ensemble $\varepsilon$, which is described by (\ref{eq:gtichiji}), is the union of all the $\varepsilon_{i,\alpha}$, the density matrix $\rho^{\prime}$ describing $\varepsilon$ should be written as the weighted sum of all the $\rho_{i,\alpha}$:
\begin{equation}
\rho^{\prime}=\sum_{i,\alpha}P_{i,\alpha}\rho_{i,\alpha },
\end{equation}
where $P_{i,\alpha}$ are suitable factors.  However, $\varepsilon$ is defined to be described by (\ref{eq:gtichiji}), such that it should be described by the density matrix (\ref{eq:matrix}).  $\rho_1(t)$ and $\rho^{\prime}$ are necessarily different, except in the case that $|\Phi_1(t)\rangle$ is a product of a vector $|S\rangle$ in the Hilbert space of  the observed system and a vector $|M\rangle$ in the Hilbert space of the measuring apparatus, i.e.,
\begin{equation}
|\Phi_1(t)\rangle =|S\rangle |M\rangle.
\end{equation}
(\ref{eq:gtichiji}) does not have this form.  Thus, the previous assumption has been shown to be false.

We must say for the above reason that both the observed system and the measuring apparatus do not have complete sets of their own, though they have no unitary interaction after $t$ and
\begin{equation}
{\rm Tr}[\rho_1(t)\hat F\hat x]={\rm Tr}[\rho_1(t)\hat x\hat F].\label{eq:FxxF}
\end{equation}
This is the manifestation of the quantum nonseparability\cite{Esp}, which implies that reading the pointer's position of the measuring apparatus and the post-selection are mutually dependent even after the interaction between the observed system and the measuring apparatus has ended.  In other words, the mixed state of the measuring apparatus, which is described by (\ref{eq:rhoF}), does not reflect its state after the unitary evolution by the Hamiltonian (\ref{eq:Hamiltonian}).  Hence, we cannot expect that the weak value $\langle\hat A\rangle_{FI}^w$ gives the expectation value of $\hat A$ for the initial state $|I\rangle$ in any situation, though they are regarded as expectation values in some papers as referred in the first section. 

We can expect that $\overline x_F$ gives the expectation value of $\hat A$ for a different state if and only if $[\hat A, \hat F]=0$.  In this case, 
\begin{equation}
\rho^{(m)}_{1F}(t)={\rm Tr}^{(s)}[(\hat 1-\frac{igt}{\hbar}\hat A\pi)\hat F|I\rangle |\phi\rangle\langle \phi|\langle I|\hat F(\hat 1+\frac{igt}{\hbar}\hat A\hat\pi)],
\end{equation}
so that $\overline x_F$ becomes
\begin{equation}
%\begin{split}
\overline x_F=x_0+gt\langle I|\hat F\hat A\hat F|I\rangle.
%&=x_0+gt\langle I|\hat F\hat A|I\rangle.
%\end{split}
\end{equation}
Thus, $\overline x_F$ corresponds to the expectation value of $\hat A$ for the state obtained after the projective measurement of $\hat F$ for $|I\rangle$.

\section{Simultaneous measurability and simultaneous readability}
We extend the discussion in the previous section to the von Neumann-type measurement\cite{von} and define the {\it simultaneous readability}, which is the separability between multiple measuring apparatuses,  to study the simultaneous measurability of multiple observables of an observed system from the viewpoint of the quantum measurement theory.

If the ensembles $M_A$ and $M_B$ of two measuring apparatuses are both separately obtained by combining all the elements of the subensembles, each of them is described by its own ket.  Then, each element of $M_A$ belongs to one of the subensembles $E_{\alpha},\ \alpha =1,2,\cdots$ described by $|a_{\alpha}\rangle$ and each element of $M_B$ belongs to one of the subensembles $E_{\beta},\ \beta =1,2,\cdots$ described by $|b_{\beta}\rangle$, such that the subensemble $\varepsilon_{\alpha,\beta}$ of the combined measuring apparatus, whose elements belong to both $E_{\alpha}$ and $E_{\beta}$, is described by the density matrix
\[
\rho_{\alpha ,\beta}=|b_{\beta}\rangle |a_{\alpha}\rangle\langle a_{\alpha}|\langle b_{\beta}|.
\]
Thus, the ensemble of the measuring apparatuses is described as the weighted sum of all the $\rho_{\alpha,\beta}$:
\begin{equation}
\rho^{\prime}=\sum_{\alpha ,\beta}P_{\alpha ,\beta}\rho_{\alpha ,\beta},\label{eq:simulrhoprime}
\end{equation}
where $P_{\alpha ,\beta}$ are suitable factors. 
Then, we define the simultaneous readability as follows.  Let $|\Psi(t)\rangle$ be the entangled state after the unitary interaction between the observed system and the measuring apparatuses and $\rho^{(m)}(t)$ be the partial density matrix only of the measuring apparatuses defined as
\begin{equation}
\rho^{(m)}(t)\equiv {\rm Tr}^{(s)}\big[\rho(t)\big],\label{eq:simulrho}
\end{equation}
where
\begin{equation}
\rho(t)=|\Psi(t)\rangle\langle\Psi(t)|.
\end{equation}
We call the observables of the different measuring apparatuses simultaneously readable if and only if %all of their expectation values which are simultaneously given by $\rho^{(m)}(t)$ agree with what are separately given by $\rho^{(m)}(t)$, respectively.  This is equivalent to the condition that
 $\rho^{(m)}(t)$ is written in the form of (\ref{eq:simulrhoprime}), that is
\begin{equation}
\rho^{(m)}(t)=\sum_{\alpha ,\beta}P_{\alpha ,\beta}\rho_{\alpha ,\beta}.\label{eq:31}
\end{equation}
If not, these observables are nonseparable, that is, both the measuring apparatuses do not have complete sets of their own.

Next, we discuss it more specifically to see the relation between the simultaneous readability and the simultaneous measurability.  Let $|I\rangle$ be the initial state of the observed system and $|\phi\rangle$ and $|\psi\rangle$ be the initial states of the measuring apparatuses of two observables $\hat A$ and $\hat B$ of the observed system, respectively.  Then, the initial state of the unified system $|\Psi (0)\rangle$ is
\begin{equation}
|\Psi(0)\rangle =|I\rangle |\phi\rangle |\psi\rangle.\label{eq:18}
\end{equation}
The initial wavefunctions of the measuring apparatuses are assumed to be
\begin{equation}
\phi (x_A)=\langle x_A|\phi\rangle =\Big({1\over \sqrt{2\pi}\sigma_A}\Big)^{1/2}\exp\Big( -{(x_A-(x_A)_0)^2\over 4\sigma_A^{\ 2}}\Big),\label{eq:phix}
\end{equation}
\begin{equation}
\psi (x_B)=\langle x_B|\psi\rangle =\Big({1\over \sqrt{2\pi}\sigma_B}\Big)^{1/2}\exp\Big( -{(x_B-(x_B)_0)^2\over 4\sigma_B^{\ 2}}\Big).\label{eq:psix}
\end{equation}
We define the interaction Hamiltonian\cite{Resch1} as
\begin{equation}
\hat H_I=g_A\hat A\hat\pi_A +g_B\hat B\hat\pi_B,\label{eq:interaction}
\end{equation}
where $g_A$ and $g_B$ are coupling constants and $\hat \pi_A$ and $\hat\pi_B$ are the momenta of the pointers.  In the same way as the previous section, we assume $\hat H_I$ to be constant and dominant over some interaction time $t$.  Then, the state after the interaction is
\begin{equation}
|\Psi(t)\rangle =\exp\Big(-\frac{i\hat H_It}{\hbar}\Big)|\Psi(0)\rangle.\label{eq:simul}
\end{equation}
If we read the position of only one pointer of these measuring apparatuses, its readout can be interpreted as the expectation value of the corresponding observable: 
\begin{equation}
\begin{split}
\overline x_A&={\rm Tr}\big[\rho(t)\hat x_A\big]\\
%&=\langle\Psi(0)|\exp \big(+\frac{i\hat H_It}{\hbar}\big)\hat x_A\exp \big(-\frac{i\hat H_It}{\hbar}\big)|\Psi(0)\rangle \\
%&=\langle\Psi(0)|\big(\hat x_A+\frac{ig_At}{\hbar}\hat A[\hat\pi_A, \hat x_A]\big)|\Psi(0)\rangle \\
%&=\langle\phi |\hat x_A|\phi\rangle+g_At\langle I|\hat A|I\rangle \\
&=(x_A)_0+g_At\langle I|\hat A|I\rangle,\label{eq:23}
\end{split}
\end{equation}
\begin{equation}
\begin{split}
\overline x_B&={\rm Tr}\big[\rho(t)\hat x_B\big]\\
%&=\langle\psi |\hat x_B|\psi\rangle+g_Bt\langle I|\hat B|I\rangle \\
&=(x_B)_0+g_Bt\langle I|\hat B|I\rangle,
\end{split}
\end{equation}
where
\begin{equation}
\rho(t)=|\Psi(t)\rangle\langle\Psi(t)|.
\end{equation}

However, both of the readouts do not, in general, give the information of the observables of the observed system simultaneously.  It is because the partial density matrix  $\rho^{(m)}(t)$ made from (\ref{eq:simul}) does not have the form (\ref{eq:31}) in general.  That is, $\hat x_A$ and $\hat x_B$ are not always simultaneous readable. On the other hand, their product $\hat x_A\hat x_B$ is an observable of the combined measuring apparatus because $[\hat x_A, \hat x_B]=0$.  We can calculate the expectation value of $\hat x_A\hat x_B$:
\begin{equation}
\begin{split}
\overline{( x_Ax_B)}&={\rm Tr}\big[\rho(t)\hat x_A\hat x_B\big]\\
%&=\langle\Psi(0)|\exp \big(+\frac{i\hat H_It}{\hbar}\big)\hat x_A\hat x_B\exp \big(-\frac{i\hat H_It}{\hbar}\big)|\Psi(0)\rangle \\
%&=\langle\Psi(0)|\big(\hat x_A\hat x_B+g_At\hat A\hat x_B+g_Bt\hat B\hat x_A+\frac{1}{2}g_Ag_Bt^2(\hat A\hat B+\hat B\hat A)\big)|\Psi(0)\rangle \\
%&=\langle\phi |\hat x_A|\phi\rangle\langle\psi |\hat x_B|\psi\rangle+g_At\langle\psi |\hat x_B|\psi\rangle\langle I|\hat A|I\rangle+g_Bt\langle\phi |\hat x_A|\phi\rangle\langle I|\hat B|I\rangle\\
%&\ \ +\frac{1}{2}g_Ag_Bt^2\langle I|(\hat A\hat B+\hat B\hat A)|I\rangle \\
&=(x_A)_0(x_B)_0+g_At(x_B)_0\langle I|\hat A|I\rangle+g_Bt(x_A)_0\langle I|\hat B|I\rangle\\
&\ \ +\frac{1}{2}g_Ag_Bt^2\langle I|(\hat A\hat B+\hat B\hat A)|I\rangle+\cdots.\label{eq:doujidame}
\end{split}
\end{equation}
This is the expectation value of the observable $(1/2)(\hat A\hat B+\hat B\hat A)$ for the initial state $|I\rangle$, when $(x_A)_0=(x_B)_0=0$ and $[\hat A, \hat B]$ is a {\it C}-number.  However, we should pay attention to the following fact 
\[
\overline{( x_Ax_B)}\ne\overline x_A\overline x_B,
\]
such that $\overline x_A$ and $\overline x_B$ cannot be separately extracted from (\ref{eq:doujidame}).

%The following discussion is almost the same as that in the previous section.  Let $\varepsilon$ be the ensemble described by the partial density matrix $\rho^{(m)}(t)$ only of the measuring apparatuses defined as

If and only if $[\hat A, \hat B]=0$, (\ref{eq:simul}) becomes
\begin{equation}
|\Psi(t)\rangle=\exp\Big(-\frac{ig_At}{\hbar}\hat A\hat\pi_A\Big)|\phi\rangle\exp\Big(-\frac{ig_Bt}{\hbar}\hat B\hat\pi_B\Big)|\psi\rangle |I\rangle,\label{eq:29}
\end{equation}
and the partial density matrix $\rho^{(m)}(t)$ is in the form of (\ref{eq:31}).  In this case, $\hat x_A$ and $\hat x_B$ are simultaneous readable and thus we can simultaneously regard both the readouts of the measuring apparatuses as the expectation values of $\hat A$ and $\hat B$, respectively.  If not, only one of three observables $\hat x_A$, $\hat x_B$ and $\hat x_A\hat x_B$ can be read to receive the information of the observed system.  Thus, we can say that $\hat A$ and $\hat B$ with the interaction Hamiltonian (\ref{eq:interaction}) are simultaneously measurable if and only if $\hat x_A$ and $\hat x_B$ are simultaneously readable.  The above discussion suggests that reading the outcomes of the measuring apparatuses mutually affect even after the unitary part of the measuring process has ended.  Although we can simultaneously read the values $\overline x_A$ and $\overline x_B$ on the measuring apparatuses, we cannot regard them as the expectation values of the noncommuting observables $\hat A$ and $\hat B$, respectively.

  If we identify $\hat B$ with the projection operator $\hat F$ of the post-selection, we can reproduce the conclusion in the second section.  When  $[\hat A,\hat F]\ne0$, $\overline{(x_Ax_F)}$ corresponds to the expectation value of the observable in the Hilbert space of the observed system; however, both $\overline x_A$ and $\overline x_F$ do not.  If we select a final state in the weak measurement, we cannot obtain the expectation value of the observed system from the readout of the measurement apparatus. 
%この節の議論で被測定系のオブザーバブル$\hat B$を，弱測定において終状態を選ぶ射影演算子$\hat F$とすれば、2節の結論を再び得ることができる。のとき、は被測定系のオブザーバブルに対応物を持つが、そのことは、が同時に被測定系のオブザーバブルに対応することを意味しない。我々が測定器から読み取れるのはeither $\hat A$ or $\hat F$である。

\section{Weak measurement without post-selection}
To study the simultaneous measurability by the weak measurement without post-selection, we expand (\ref{eq:simul}) to the first order of $g_A$ and $g_B$:
\begin{equation}
%\begin{split}
|\Psi(t)\rangle =|I\rangle |\phi\rangle |\psi\rangle -\frac{ig_At}{\hbar}\hat A|I\rangle\hat \pi_A|\phi\rangle |\psi\rangle-\frac{ig_Bt}{\hbar}\hat B|I\rangle |\phi\rangle\hat \pi_B|\psi\rangle.\label{eq:weak}%\\
%&=\big(1-\frac{ig_At}{\hbar}\hat A\hat\pi_A\big)|\phi\rangle\big(1-\frac{ig_Bt}{\hbar}\hat B\hat\pi_B\big)|\psi\rangle |I\rangle.
%\end{split}
\end{equation}
Then, its partial density matrix $\rho^{(m)}(t)$ satisfies (\ref{eq:31}) up to the first order of the couplings.  Nevertheless, we do not think that this fact shows the simultaneous measurability of $\hat A$ and $\hat B$.  Suppose that two observables are measured in turns and their expectation values are obtained after the series of measurements.  Then, is it appropriate to regard them as simultaneously measurable?  It shows only a similar situation.  The commutator of $\hat A$ and $\hat B$ appears in the second order of the couplings; thus, we cannot discuss the simultaneous measurability in the first order of the couplings.  

Up to the second order of the couplings, $|\Psi(t)\rangle$ becomes
\begin{equation}
|\Psi(t)\rangle=\big(1-iH_It-\frac{1}{2}H_I^{\ 2}t^2\big)|I\rangle,
\end{equation}
whose partial density matrix $\rho^{(m)}(t)$ does not satisfy (\ref{eq:31}) if $[\hat A, \hat B]\ne 0$.  Thus, $\hat x_A$ and $\hat x_B$ are not simultaneously readable or $\hat A$ and $\hat B$ are not simultaneously measurable.  On the other hand, with $(x_A)_0=(x_B)_0=0$,
\begin{equation}
\overline{(x_Ax_B)}
=\frac{g_Ag_Bt^2}{2}\langle I|(\hat A\hat B+\hat B\hat A)|I\rangle,\label{eq:secondorder})
\end{equation}
which agrees with (\ref{eq:doujidame}).  %Lundeen {\it et al.}\cite{Lun2} have suggested the method to determine the density matrix with the aid of the second order weak measurement without the post-selection.  As previously noted, $\hat x_A\hat x_B$ is an observable of the combined measuring apparatus, because $[\hat x_A, \hat x_B]=0$.  Thus, we have no reason to deny their discussion.  %Provided that (\ref{eq:secondorder}) does not give simultaneously the separate expectation values of $\hat A$ and $\hat B$.

\section{Measuring a different observable for eigenstates}

We consider the spin of an electron with the appropriate normalization.  $x$-component $\hat \sigma_x$ and $z$-component $\hat \sigma_z$ of the spin are not simultaneously measurable or their corresponding observables of the measuring apparatuses are not simultaneously readable.  It is not obvious, however, whether the following question has the same answer\cite{Heisenberg}\cite{Ozawa2}\cite{Ozawa1}:  An electron is prepared in the eigenstate of $\hat\sigma_z$ at time $0$ and its $\hat \sigma_x$ will be measured for this state.  Then, can we know both the expectation values of $\hat\sigma_z$ and $\hat\sigma_x$ at time $0$?  

Let $|\phi\rangle$ be the initial state of the measuring apparatus of $\hat\sigma_x$ and $|I\rangle$ be the initial state of the observed system.  The state of the unified system after the interaction is
\begin{equation}
|\Phi(t)\rangle=\exp\Big(-\frac{igt}{\hbar}\hat\sigma_x\hat\pi\Big)|\Phi(0)\rangle\label{eq:epr}
\end{equation}
and its density matrix is
\begin{equation}
\rho(t)=|\Phi(t)\rangle\langle\Phi(t)|,\label{eq:eprmatrix}
\end{equation}
where
\begin{equation}
|\Phi(0)\rangle=|I\rangle |\phi\rangle.\label{eq:initialstate}
\end{equation}
We prepare $|\uparrow\rangle$, which is the eigenstate of $\hat\sigma_z$ with an eigenvalue $+1$, as the initial state of the observed system.  

In the second section, we considered the weak measurement, and almost the same discussion can be applied to the von Neumann-type measurement.  If we take the trace only in the Hilbert space of the measuring apparatus,  (\ref{eq:eprmatrix}) becomes the proper density matrix of the observed system, and vice versa.  Nevertheless, the observed system and the measuring apparatus are nonseparable, so we must lose the information of the observed system at time $t$ if we read the value on the measuring apparatus.  On the other hand, because the evolution from time $0$ to $t$ is unitary, we can calculate the state at time $t$ if we know that at time $0$.  Thus, we  must conclude that we have no information for the observed state even at time $0$ after we read the outcome of the measuring apparatus.

However, we prepared the eigenstate $|\uparrow\rangle$ as the initial state.   We need to reconsider {\it the eigenstate} to convince ourselves that there is no contradiction between these statements.   We will necessarily obtain $+1$ if we measure {\it only} $\hat\sigma_z$ for $|\uparrow\rangle$.  We should not express this fact  by the statement ``The $z$-component of the spin of $|\uparrow\rangle$ is $+1$".  The eigenstate should be understood contextually.  We have nothing to say about the $z$-component of the spin of the initial state if we read its $x$-component.  Thus, our answer to the previous question is NO.  We cannot know the different components of the spin simultaneously even if the eigenstate of one component of the spin is prepared as the initial state.  The above discussion may make us remember the delayed-choice experiment\cite{Wheeler}\cite{Tang}, which seems very strange but is the proper consequence of quantum mechanics.

\section{Simultaneous measurement of an EPR-correlated electron}
We consider a pair of electrons that have the EPR correlation\cite{EPR}\cite{Bohm} and their total spin of $0$.  The operator $\hat C_z$, which measures the correlation, is defined as
\begin{equation}
\hat C_z\equiv (\hat\sigma_z)_1(\hat\sigma_z)_2,
\end{equation}
where $(\hat\sigma_z)_{1}$ and $(\hat\sigma_z)_{2}$ are the $z$-components of the spins of the electrons 1 and 2, respectively.  We prepare $|-1\rangle$, which is the eigenstate of $\hat C_z$ with an eigenvalue $-1$, as the initial state $|I\rangle$ of the observed system, i.e. the combined system of these electrons.  The initial state of the unified system of the observed system and the measuring apparatuses is
\begin{equation}
|\Psi(0)\rangle=|I\rangle |\phi\rangle |\psi\rangle,
\end{equation}
where $|\phi\rangle$ and $|\psi\rangle$ are the initial states of the measuring apparatuses of the electrons 1 and 2, respectively.  Their wavefunctions are given in (\ref{eq:phix}) and (\ref{eq:psix}).  The interaction Hamiltonian is
\begin{equation}
\hat H_I=g_A(\hat\sigma_x)_1\hat\pi_A+g_B(\hat\sigma_z)_2\hat\pi_B.
\end{equation}
Then, the state after the unitary evolution is
\begin{equation}
\begin{split}
|\Psi(t)\rangle &=\exp\Big(-\frac{i\hat H_It}{\hbar}\Big)|\Psi(0)\rangle\\
& =\exp\Big(-\frac{i\hat H_{I1}t}{\hbar}\Big)|\phi\rangle\exp\Big(-\frac{i\hat H_{I2}t}{\hbar}\Big)|\psi\rangle |I\rangle,\label{eq:eprt}
\end{split}
\end{equation}
where
\[
\hat H_{I1}=g_A(\hat\sigma_x)_1\hat\pi_A,
\]
\[
\hat H_{I2}=g_B(\hat\sigma_z)_2\hat\pi_B.
\]
Because the partial density matrix $\rho^{(m)}(t)$, which is only for the measuring apparatuses made from (\ref{eq:eprt}), has the form of (\ref{eq:31}), $\hat x_A$ and $\hat x_B$ are simultaneously readable.  Thus, we can expect that both the measuring apparatuses give the proper outcomes simultaneously.  

Then, it remains to be seen whether (\ref{eq:eprt}) implies the simultaneous measurability of different components of the spin of a single electron if we take account of the EPR correlation of the electrons.  The fact that they have complete negative correlation is shown in the following equation:
\begin{equation}
|I\rangle=\frac{1-\hat C_z}{2}|I\rangle.\label{eq:machigai}
\end{equation}
%$(\hat \sigma_z)^2=1$ and 
%\[
%(\hat\sigma_z)_2\frac{1-\hat C}{2}=(\hat\sigma_z)_1\frac{\hat C-1}{2}=-(\hat\sigma_z)_1\frac{1-\hat C}{2},
%\]
If we use this equation, (\ref{eq:eprt}) is rewritten as
\begin{equation}
\begin{split}
|\Psi(t)\rangle&=\exp\Big(-\frac{i\hat H_{I1}t}{\hbar}\Big)|\phi\rangle\exp\Big(-\frac{i\hat H_{I2}t}{\hbar}\Big)|\psi\rangle\frac{1-\hat C_z}{2} |I\rangle\\
&=\exp\Big(-\frac{i\hat H_{I1}t}{\hbar}\Big)|\phi\rangle\exp\Big(+\frac{i\hat H_{I2}^{\prime}t}{\hbar}\Big)|\psi\rangle |I\rangle,\label{eq:dekiru} 
\end{split}
\end{equation}
where
\[
\hat H_{I2}^{\prime}=g_B(\hat\sigma_z)_1\hat\pi_B.
\]
If (\ref{eq:dekiru}) is justified, we can insist that different components of the spin of a single electron are simultaneously measurable, because the partial density matrix $\rho^{(m)}(t)$ made from (\ref{eq:dekiru}) once again has the form of (\ref{eq:31}) and $\hat x_A$ and $\hat x_B$ are simultaneously readable.

However, (\ref{eq:dekiru}) is not justified because of the following reason.   It is shown in (\ref{eq:eprt}) that the electron pair and each of the measuring apparatuses are nonseparable.  Though we have prepared $|-1\rangle$ as the initial state, this only means that we will necessarily obtain $-1$ if we only measure $\hat C_z$ for $|-1\rangle$ as discussed in the previous section. Conversely, if we read the outcome of the measuring apparatus of $(\hat\sigma_x)_1$,  we will receive no information about the correlation of our electron pair at time $t$ and, by extension,  at time $0$.  (\ref{eq:machigai}) is not a definition or an identity but an equation that holds only contextually.  This equation cannot be used in the context in this section.  We can measure any component of the spin of each electron of the EPR-correlated pair but cannot regard it as the measurement of a pair of components of the spin of a {\it single} electron.

\section{Concluding remarks}

In our discussion, the density matrix of the unified system of the observed system and the measuring apparatuses has played an essential role.  This finding reflects that the nonseparability and the nonunitary change, i.e., the reduction of the wavepacket or its replacement, are indispensable in the quantum measurement theory. We have not discussed the problem of the von Neumann chain\cite{Shimizu} about the extent where the quantum measurement theory treats the reading of the measuring apparatus; thus our word {\it read} contains some ambiguity in its meaning.  Nevertheless, we have shown that reading the outcome of the measuring apparatus, which may be a series of actions, brings the inevitable effect on the unified system, though it is after the unitary interaction between the observed system and the measuring apparatuses.  By this observation, we have suggested the simultaneous readability, which is the separability between the measuring apparatuses.  This is the necessary and sufficient condition of the simultaneous measurability if the interaction Hamiltonian between the observed system and the measuring apparatuses is (\ref{eq:interaction}).  To understand the simultaneous measurability, it is not sufficient to study the unitary evolution of the unified system.  From the viewpoint of the quantum measurement theory, it is in the nonunitary change that the essence of the simultaneous measurability exists.   

In this context, the discussion in the second section is valid no matter how weak the measurement is.   Starting with (\ref{eq:full}), we obtain 
\begin{equation}
\begin{split}
{\rm Tr}\big[\rho(t)\hat F\big]&={\rm Tr}\big[\exp\Big(-\frac{i\hat H_It}{\hbar}\Big)|\Phi(0)\rangle\langle\Phi(0)|\exp\Big(+\frac{i\hat H_It}{\hbar}\Big)\hat F\big]\\
&=\langle I|\hat F|I\rangle
\end{split}
\end{equation}
if $[\hat A,\hat F]$ is a $c$-number.  This equation shows that we can receive information about the initial state of the observed system even after its unitary interaction with the measuring apparatus, which is not necessarily weak, has ended.  It is the nonunitary change accompanied with reading the outcome of the measuring apparatus that hides the information of the initial state.  We well convince this fact to ourselves if we reinterpret (\ref{eq:29}) as the state after a series of interactions between the observed system and the measuring apparatuses, that is, the interactions of $\hat B$ between $0$ and $t$ and $\hat A$ between $t$ and $2t$.  Then, even if $[\hat A,\hat B]\ne 0$, the unified state at time $2t$ is
\begin{equation}
|\Psi(2t)\rangle=\exp\Big(-\frac{ig_At}{\hbar}\hat A\hat\pi_A\Big)|\phi\rangle\exp\Big(-\frac{ig_Bt}{\hbar}\hat B\hat\pi_B\Big)|\psi\rangle |I\rangle.\label{eq:last}
\end{equation}
If we read only $\overline x_A$, we obtain the same result as (\ref{eq:23}).  The unitary part of the measurement of $\hat B$ has no effect on this result.

Moreover, it is worth noting that $\hat x_A$ and $\hat x_B$ are simultaneously readable if the unified system is expressed by (\ref{eq:last}).  We can know the expectation values of $\hat B$ at time $0$ and $\hat A$ at $t$ from those of $\hat x_B$ and $\hat x_A$, respectively.  However, the expectation value of $\hat A$ at time $t$ after the reading of the outcome of $\hat B$ does not agree with (\ref{eq:23}) in general.  We will obtain the proper expectation value of $\hat A$ according to the readout of the measuring apparatus of $\hat B$; thus we cannot know the expectation values of $\hat A$ and $\hat B$ at time $0$ simultaneously.  On the other hand, as discussed in the fifth section, we cannot know the expectation values of $\hat A$ and $\hat B$ at time $t$ simultaneously because the observed system and the measuring apparatus of $\hat A$ are nonseparable.  Thus, we conclude that the noncommuting observables $\hat A$ and $\hat B$ are not simultaneously measurable, though  $\hat x_A$ and $\hat x_B$ are simultaneously readable in this case.

As noted in the above paragraph, we have shown in the fifth section that eigenstates should be understood contextually.  Thus, Bell's inequality\cite{Bell} is not concluded even though a quantum state is assumed to be a simultaneous eigenstate of multiple noncommuting observables.  Nevertheless, it does not mean that we can obtain the expectation values of the noncommuting observables from such a state simultaneously.  Thus, we do not need to describe the state as the simultaneous eigenstate of noncommuting observables; the quantum mechanics still stay complete for describing our possible knowledge about the quantum state.  %In addition, it would not support any local hidden variable theories.  Bell-nonlocality has not vanished but has shifted between the observed system and the measuring apparatus or between the multiple measuring apparatuses.  


\begin{thebibliography}{99}
\bibitem{Esp}B. d'Espagnat, ``Conceptual Foundations of Quantum Mechanics,"2nd Ed., W. A. Benjamin, Menlo Park, California, 1976
\bibitem{EPR}A. Einstein, B. Podolsky and N. Rosen, {\it Phys. Rev.} {\bf 47}, 777 (1935)
\bibitem{Bohm}D. Bohm, ``Quantum Theory,"Prentice-Hall, Englewood Cliffs, 1951
\bibitem{Aha2}Y. Aharonov, D. Z. Albert and L. Vaidman, {\it Phys. Rev. Lett.} {\bf 60}, 1351 (1988)
\bibitem{Aha25}Y. Aharonov and L. Vaidman, {\it Phys. Rev. A} {\bf 41}, 11 (1990)
\bibitem{Lun1}J. S. Lundeen, B. Sutherland, A. Patel C. Stewart and C. Bamber, {\it Nature} {\bf 474}, 188 (2011)
\bibitem{Mal}M. Malik, M. Mirhoseinni, M. P. J. Lavery, J.Leach, M. J. Padgett and R. W. Boyd, {\it Nature Communication} {\bf 5}, 3115 (2014)
\bibitem{Sal}J. Z. Salvail, M. Agnew, A. S. Johnson, E. Bolduc, J. Leach and R. W. Boyd, {\it Nature Photonics} {\bf 7}, 316 (2013)
\bibitem{Aha3}Y. Aharonov and A. Botero, {\it Phys. Rev. A} {\bf 72}, 052111 (2005)
\bibitem{Kocsis}S. Kocsis, B. Braverman, S. Ravets, M. J. Stevens, R. P. Mirin, L. K. Shalm, and A. M. Steinberg, {\it Science} {\bf 332}, 1170 (2011).
\bibitem{Wise1}H. M. Wiseman, {\it New J. Phys.} {\bf 9}, 165 (2007).
\bibitem{Hosoya2}A. Hosoya and Y. Shikano, {\it J. Phys. A: Math. Theor.} {\bf 43}, 385307 (2010)
\bibitem{Aha4}Y. Aharonov, A. Botero, S. Popescu, B. Reznik and J. Tollaksen, {\it Phys. Lett.} {\bf 301}, 130 (2002)
\bibitem{Mochi}R. Mochizuki, {\it Prog. Theor. Exp. Phys.} 043A02 (2014)
\bibitem{Dre1}J. Dressel and A. N. Jordan, {\it Phys. Rev. A} {\bf 85}, 012107 (2012) 
\bibitem{von}J. von Neumann, ``Die Mathematische Grundlagen der Quantenmechanik," Springer Verlag, Verlin, 1932
\bibitem{Heisenberg}W. Heisenberg, ``The Physical Principles of the Quantum Theory,"University of Chicago Press, Chicago, 1930
\bibitem{Ozawa2}M. Ozawa, {\it Found. Phys.} {\bf 41}, 592 (2011)
\bibitem{Ozawa1}M. Ozawa, {\it AIP Conf. Proc.} {\bf 1363}, 53 (2011)
\bibitem{Resch1}K. J. Resch and A. M. Steinberg, {\it Phys. Rev. Lett.} {\bf 92}, 130402 (2004)
\bibitem{Wheeler}J. A. Wheeler,  in ``Mathematical Foundations of Quantum Theory (eds A. R. Marlow),"Academic Press, 1978
\bibitem{Tang}J.-S. Tang, Y.-L. Li, X.-Y. Xu, G.-Y. Xiang, C.-F. Li and G.-C. Guo, {\it Nature Photonics} {\bf 6}, 600 (2012)
\bibitem{Shimizu}K. Koshino and A. Shimizu, {\it Phys. Rep.} {\bf 412}, 191 (2005)
\bibitem{Bell}J. S. Bell, {\it Physics} {\bf 1}, 195 (1964)

%\bibitem{Koshino}K. Koshino and A. Shimizu, {\it Phys. Rep.} {\bf 412}, 191 (2005)
%\bibitem{Hosoya1}A. Hosoya and M. Koga, {\it J. Phys. A} {\bf 44}, 415303 (2011)
%\bibitem{Tresser1}C Tresser, {\it Eur. Phys. J.} {\bf D58}, 385 (2010)
%\bibitem{Hardy}L. Hardy, {\it Phys. Rev. Lett.} {\bf 68}, 2981 (1992)
%\bibitem{Irvine}W. T. M. Irvine, J. F. Hodelin, C. Simon and D. Bouwmeester, {\it Phys. Rev. Lett.} {\bf 95}, 030401 (2005)
%\bibitem{Leggett}A. J. Leggett, {\it Phys. Rev. Lett.} {\bf 62}, 2325 (1989)
%\bibitem{Duck}I. M. Duck, P. M. Stevenson and E. C. G. Sudarshan, {\it Phys. Rev. D} {\bf 40}, 2112 (1989)
%\bibitem{Dressel}J. Dressel and A. N. Jordan, {\it Phys. Rev. A} {\bf 85}, 012107 (2012)
%\bibitem{Svozil}K. Svozil, ``Quantum Logic," Springer-Verlag, Singapore City (1998)
%\bibitem{Maeda}S. Maeda, ``Lattice Theory and Quantum Logic," Maki-Shoten, Tokyo (1980)
%\bibitem{Aha1}Y. Aharonov, P. G. Bergmann and J. L. Lebowitz, {\it Phys. Rev. }{\bf 134}, B1430 (1964)
%\bibitem{Aha6}Y. Aharonov and L. Vaidman, {\it J. Phys. A: Math. Gen.} {\bf 24}, 2315 (1991)
%\bibitem{Vaidman}L. Vaidman, {\it Found. Phys.} {\bf 26}, 895 (1996)
%\bibitem{swift}R. Swift and R. Wright, {\it J. Math. Phys.} {\bf 21}, 77 (1979)
%\bibitem{Lun2}J. S. Lundeen and C. Bamber,  {\it Phys. Rev. Lett.} {\bf 108} 070402 (2012)
%\bibitem{Resch2}K. J. Resch, J. S. Lundeen and A. M. Steinberg, {\it Phys. Lett. A} {\bf 324}, 125 (2004)
%\bibitem{Lundeen}J. S. Lundeen and A. M. Steinberg, {\it Phys. Rev. Lett.} {\bf 102}, 020404 (2009)
%\bibitem{Yokota}K. Yokota, T, Yamamoto, M. Koashi and N. Imoto, {\it New J. Phys.} {\bf 11}, 033011 (2009)


\end{thebibliography}
\end{document}